# Analytical method to determine flexoelectric coupling coefficient at nanoscale

Hao Zhou,[1,2] Yongmao Pei,[1,a)] Jiawang Hong[3,4,a)] and Daining Fang[1,3]

[1] *State Key Laboratory for Turbulence and Complex Systems, College of Engineering, Peking University, Beijing 100871, China*
[2] *Beijing Institute of Spacecraft System Engineering, Beijing 100094, China*
[3] *Institute of Advanced Structure Technology, Beijing Institute of Technology, Beijing 100081, China*
[4] *Department of Applied Mechanics, Beijing Institute of Technology, Beijing 100081, China*

Flexoelectricity is defined as the coupling between strain gradient and polarization, which is expected to be remarkable at nanoscale. However, measuring the flexoelectricity at nanoscale is challenging. In the present work, an analytical method for measuring the flexoelectric coupling coefficient based on nanocompression technique is proposed. It is found that the flexoelectricity can induce stiffness softening of dielectric nano-cone-frustum. This phenomenon becomes more significant when the sample size decreases or the half cone angle increases. This method avoids measuring the electric polarization or current at nanoscale with dynamical loading, which can be beneficial to the flexoelectric measurement at nanoscale and design of flexoelectric nanodevices.

[a)] Authors to whom correspondence should be addressed. Electronic addresses: peiym@pku.edu.cn and hongjw04@gmail.com.

Flexoelectricity is an electromechanical coupling effect between strain/stress gradient and electric polarization (direct effect).[1-6] Different from piezoelectricity existing only in 20 crystal point groups without center of symmetry, flexoelectricity occurs in all 32 crystal point groups because of its inversion symmetry broken by strain/stress gradients. Flexoelectricity can affect various materials properties, such as ferroelectric domain configuration,[7,8] dead layer effect,[9] critical thickness for ferroelectricity,[10] imprint behavior,[11,12] size effect of stiffness,[13,14] and electric field dependence of stiffness,[15,16] etc. It can also be employed to improve materials properties and create metamaterials and new techniques, such as enhanced piezoelectricity,[17] piezoelectric devices with nonpiezoelectric materials,[18] and mechanical writing of polarization.[19,20]

The method of measuring and calculating flexoelectricity is one of the most concerned topics in this research field.[1-5,21-27] At present, there are mainly two methods to measure the flexoelectric coefficient, i.e. the beam bending method[1-3,22,28] and the compression method.[1-3] The beam bending method is to measure the effective transverse flexoelectric coefficient, whereas the compression method is to measure the effective longitudinal flexoelectric coefficient. In these two methods, both quasi-static and low frequency dynamic techniques have been employed. The mechanical bending or compression load is applied to the samples by mechanical testing machine or electro-magnetic actuator, whereas the displacement and electric charge are monitored by the strain gages and the charge amplifier or electrometer. The samples in these measurements are in the millimeter or submillimeter scale. In addition, the phonon spectra may provide information on the coupled action of the static and dynamic bulk flexoelectricity.[2,5,29] However, it's difficult to distinguish the static flexoelectricity from the dynamic flexoelectricity.

So far, it's very challenging to measure full flexoelectric coefficients at nanoscale, in which flexoelectricity may have significant effect on the materials properties.

In recent years, nanoindentation technique has been widely employed to investigate the nanoscale mechanical properties.[15-16,31-32] In order to obtain the uniaxial stress-strain relationship and mechanical properties of nanomaterials, focused ion beam technique is used to fabricate nanopillar samples, and in-situ nanoindentation instrument equipped with a flat-ended indenter tip is used to conduct the nanocompression test.[32-34] This technique is excellent in investigating the uniaxial deformation behavior of materials at nano- and microscale. By fabricating the variable cross-section nanopillars and using this nanoindentation technique, it is possible to induce stress gradient in nanopillars and measure the modified stress-strain relationship due to the stress gradient. In the present work, we developed a phenomenological method to investigate the influence of flexoelectricity on the stiffness of nanopillars measured by nanocompression technique. An analytical approach to determine the flexoelectric coupling coefficient at nanoscale is proposed. The stiffness softening of nano-cone-frustum due to the flexoelectricity is predicted. This softening becomes significant when the size of the sample decreases or the half cone angle increases. By measuring this softness, we could obtain the longitudinal flexoelectric properties of nanomaterials based on the analytical model developed in this work. This method avoids measuring the electric polarization or current at nanoscale with dynamical loading, which will simplify the setup of flexoelectric measurement.

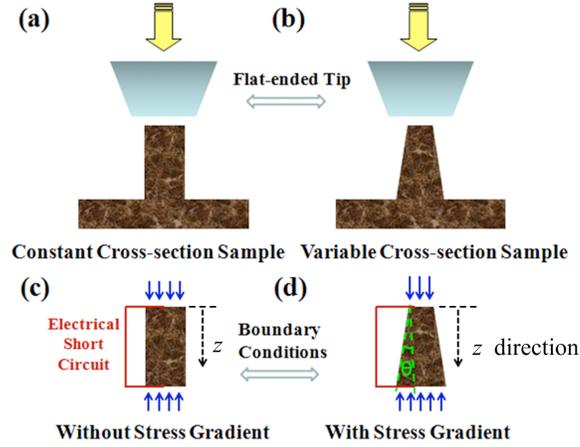

**Figure 1.** Schematic diagram of nanocompression testing. (a,c) Constant cross-section pillar sample. (b,d) Variable cross-section pillar sample.

To study the stress gradient effect, two kinds of samples with different shapes are designed, as shown in Fig. 1(a,b). Nanocompression causes homogeneous stress in the constant cross-section pillar, whereas it induces stress gradient in the longitudinal direction of the variable cross-section pillar. Since both stress and stress gradient are in the longitudinal direction (z direction shown in Fig. 1(c,d)) of the pillars, a one dimensional model is employed to study this problem. The free energy density of the system can be expressed as[35-37]

$$G(z) = \frac{1}{2}\chi^{-1}P(z)^2 + \frac{1}{2}s\sigma(z)^2 - \frac{1}{2}f_\sigma\left(P(z)\frac{d\sigma(z)}{dz} - \sigma(z)\frac{dP(z)}{dz}\right) \\ + \frac{1}{2}g\left(\frac{dP(z)}{dz}\right)^2 - P(z)E(z) - \sigma(z)\varepsilon(z)$$

(1)

where $\chi$ is the dielectric susceptibility; $P$ is the electric polarization; $s$ is the elastic compliance coefficient; $\sigma$ is the stress; $f_\sigma$ is the flexoelectric coupling coefficient, which describes the coupling between stress gradient and polarization; It is different from the flexocoupling/flexovoltage coefficient $f_\varepsilon$, which describes the coupling between strain gradient and polarization. $E$ is the electric field; $\varepsilon$ is the strain. $z$ is coordinate in space. The terms $\frac{1}{2}g\left(\frac{dP(z)}{dz}\right)^2$ has negligible effect if polarization changes slowly in the variable cross-section pillar and we will discard it to simplify the derivation.

Now that the energy density contains gradient terms, minimization of the potential of the sample as a whole (i.e. application of the Euler equations $G_X - \frac{d}{dz}G_{X'} = 0$, where $X$ stands for $P$ or $\sigma$) leads to the high order electromechanical constitutive equations:

$$\chi^{-1}P(z) - f_\sigma \frac{d\sigma(z)}{dz} = E(z), \tag{2}$$

$$s\sigma(z) + f_\sigma \frac{dP(z)}{dz} = \varepsilon(z). \tag{3}$$

Here, we consider the nanocompression process with electrical short circuit boundary condition, as shown in Fig. 1 (c,d). This identifies with the conductive property of the sample surface due to the gold-plating treatment before the nanocompression in-situ scanning electron microscopy test. In this case, the stress field $\sigma(z)$ can be easily obtained according to the one dimensional force balance equation, i.e.

$$F = \sigma(z)\cdot A(z) = \sigma(0)\cdot A(0), \tag{4}$$

where $F$ is the nanocompression force. $A(z)$ is the area of the cross-section at $z$ position of the pillar. The electrical short circuit boundary condition indicates that the electrical potential between the surface and bottom of the pillar is equal, i.e. no electric field is applied to the pillar

$$E(z) = \nabla_z \phi = \frac{\phi(H) - \phi(0)}{H} = 0, \tag{5}$$

where $\phi$ is the electrical potential; $H$ is the height of the pillar. Then, the constitutive equations can be expressed as

$$P(z) = \chi f_\sigma \frac{d\sigma(z)}{dz}, \tag{6}$$

$$\varepsilon(z) = s\sigma(z) + \chi f_\sigma^2 \frac{d^2\sigma(z)}{dz^2}. \tag{7}$$

It can be seen from Eq. (6) that mechanical stress gradient induces polarization due to flexoelectricity. From Eq. (7), we note that the mechanical strain not only results from the stress but also from the second order derivative of the stress due to the flexoelectric effect. If the second order derivative of stress posses the same (opposite) sign as the stress, the obtained stress vs. strain curve will show mechanical softening (stiffing). However, this effect will disappear in constant cross-section pillar. Therefore, the flexoelectric coupling coefficient can be determined by measuring the stiffness changes between variable cross-section and constant cross-section pillars.

The nanocompression displacement $h$ is equal to the integral of strain $\varepsilon(z)$ along the height direction of the pillar, as follows:

$$h = \int_0^H \varepsilon(z) \, dz$$

$$= \int_0^H \left[ s\sigma(0) A(0) A(z)^{-1} + f_\sigma^2 \chi \frac{d^2\sigma(z)}{dz^2} \right] dz \quad , \quad (8)$$

$$= \sigma(0) A(0) \int_0^H \left[ sA(z)^{-1} - f_\sigma^2 \chi A(z)^{-2} \frac{d^2 A(z)}{dz^2} + 2 f_\sigma^2 \chi A(z)^{-3} \left( \frac{dA(z)}{dz} \right)^2 \right] dz$$

For the column pillars, the area of the cross-section is constant:

$$A_0(z) = A_0(0) = \pi r_0^2(0). \quad (9)$$

The nanocompression stiffness is

$$C_0 = \frac{dF_0}{dh_0} = \frac{F_0}{h_0} = \frac{A_0(0)}{sH_0}. \quad (10)$$

It depends on the elastic constant and the geometric parameters of the nanopillars.

For the cone-frustum pillars, the area of the cross-section is variable:

$$A_1(z) = A_1(0) \left( 1 + \frac{\tan\theta}{r_1(0)} z \right)^2. \quad (11)$$

The nanocompression stiffness is

$$C_1 = \frac{dF_1}{dh_1} = \frac{F_1}{h_1} = \frac{A_1(0)}{\dfrac{sr_1(0)}{\tan\theta} \ln\left( \dfrac{\tan\theta}{r_1(0)} H_1 + 1 \right) - \dfrac{\chi f_\sigma^2 \tan\theta}{r_1(0)} \left[ \left( 1 + \dfrac{\tan\theta}{r_1(0)} H_1 \right)^{-2} - 1 \right]}. \quad (12)$$

It depends on not only the elastic constant and the geometric parameters, but also the flexoelectric coupling coefficient and the dielectric susceptibility of the nanopillars.

Combining Eqs. (10) and (12), an analytical characterization method to determine the flexoelectric coupling coefficient is obtained:

$$f_\sigma = \pm \sqrt{\left(\frac{A_0(0)r_1(0)}{C_0 H_0 \tan\theta}\ln\left(\frac{\tan\theta}{r_1(0)}H_1+1\right)-\frac{A_1(0)}{C_1}\right)\bigg/\left\{\frac{\chi\tan\theta}{r_1(0)}\left[\left(1+\frac{\tan\theta}{r_1(0)}H_1\right)^{-2}-1\right]\right\}}. \quad (13)$$

In the following, the dependence of nanocompression behavior on the material and geometrical parameters of the samples will be presented via numerical calculations. The default value of the physical and geometrical parameters are as follows:[1,3,7,32-34]

$s=5\times10^{-12}$ m$^2$N$^{-1}$, $\chi=2000\varepsilon_0$, $r_1(0)=0.1\times10^{-6}$ m, $r_0(0)=0.118\times10^{-6}$ m $A_1(0)=3.14\times r_1^2(0)$, $\theta=10°$, $H_1=0.5\times10^{-6}$ m, $H_0=H_1$.

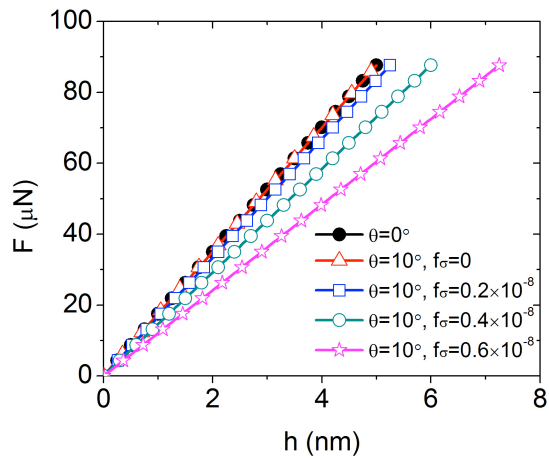

**Figure 2.** Nanocompression force-displacement curves of the column and cone-frustum samples with various flexoelectric coupling coefficient.

Fig. 2 shows the force-displacement curves of the nanocompression on the column and cone-frustum samples. For the column sample ($\theta=0°$), the force-displacement curve is independent on the flexoelectric coupling coefficient $f_\sigma$, as indicated in Eq. (10). However, for the cone-frustum samples ($\theta=10°$), the force-displacement curve is dependent on the flexoelectric coupling coefficient $f_\sigma$. The flexoelectric coupling coefficient $f_\sigma$ can be determined by the slope of linear F-h curve, as indicated in Eq. (12).

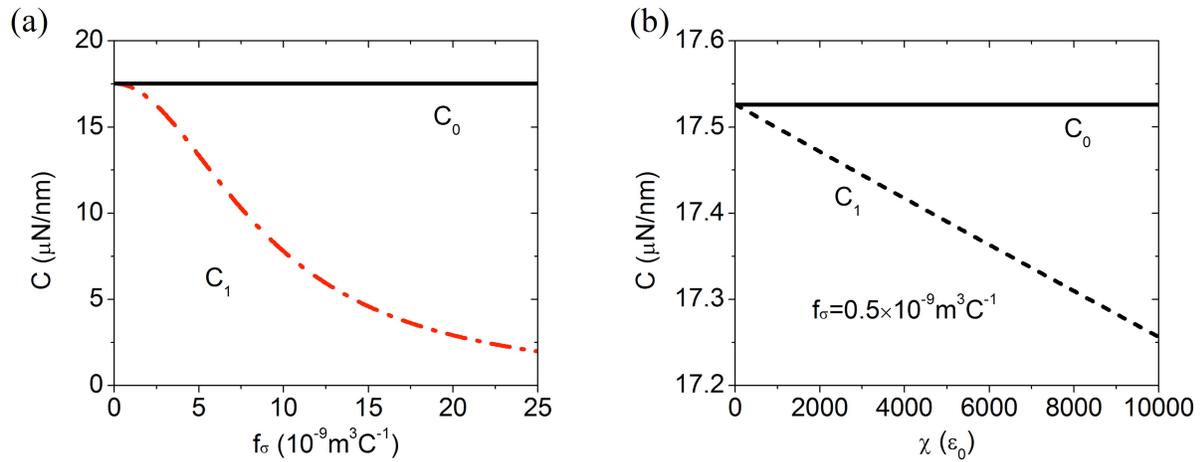

**Figure 3.** Stiffness softening phenomena. (a) Dependence of flexoelectric coupling coefficient. (b) Dependence of dielectric susceptibility.

The relationship between the nanocompression stiffness and the flexoelectric coupling coefficient or the dielectric susceptibility is shown in Fig. 3. The stiffness $C_1$ of cone-frustum sample decreases by about 80% when the flexoelectric coupling coefficient $f_\sigma$ increases by 5 times (from $5\times10^{-9}$ to $25\times10^{-9} m^3 C^{-1}$). However, the stiffness of cone-frustum sample $C_1$ decreases by about 1% when the dielectric susceptibility $\chi$ increases by 5 times (from 2000 to 10000). That is to say, the stiffness $C_1$ is sensitive to the flexoelectric coupling coefficient $f_\sigma$, but not sensitive to the dielectric susceptibility $\chi$. Therefore, the flexoelectric coupling coefficient $f_\sigma$ can be determined by an accurate measurement of the stiffness $C_1$ with an estimation of the dielectric susceptibility $\chi$. This is beneficial to the measurement at nanoscale.

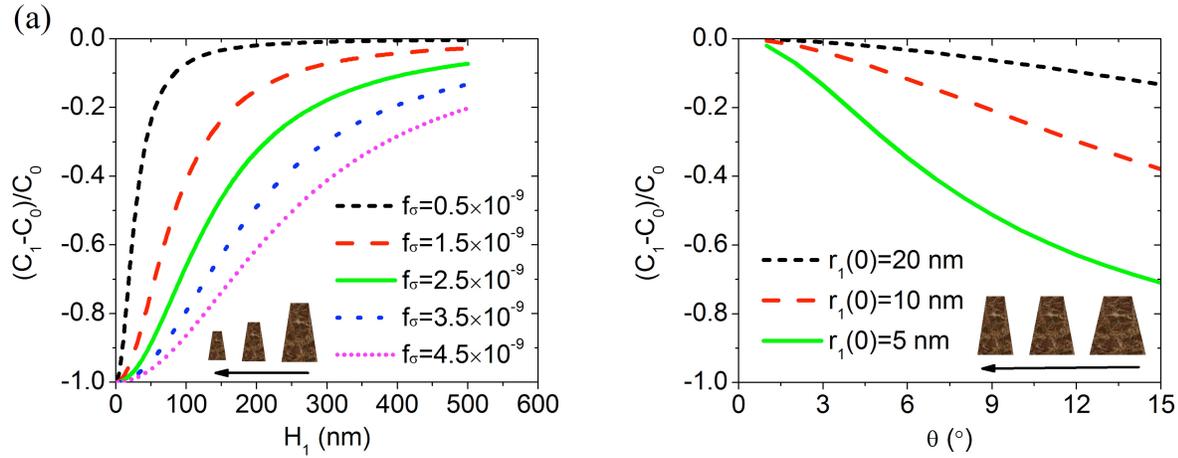

**Figure 4.** Stiffness softening phenomena. (a) Scaling effect; (b) Variable half cone angle, but constant height and top surface area.

The relationship between the stiffness reduction and the dimensional parameters are shown in Fig. 4. When the cone-frustum sample scales to smaller size, the flexoelectricity induced reduction in stiffness becomes more pronounced as shown in Fig. 4(a). When the size of the samples approaches zero, the stiffness change approaches -1, which means the stiffness of the cone-frustum sample approaches zero. The dimension range that the flexoelectricity can induce significant reduction in stiffness depends on the flexoelectric coupling coefficient. For a small flexoelectric coupling coefficient, such as $f_\sigma = 0.5 \times 10^{-9}\,\mathrm{m^3 C^{-1}}$, the reduction can be less than 5% when the height is larger than 150 nm. However, the reduction can be up to 20% when the height is 500 nm in the case of $f_\sigma = 4.5\times10^{-9}\,\mathrm{m^3 C^{-1}}$. That is to say, the smaller the scale, the larger the stress gradient and the larger the stiffness reduction induced by flexoelectricity is. In Fig. 4(b), we change the half cone angle of the cone-frustum sample with constant the height and top surface (the top radius is 5, 10 or 20 nm). The stiffness reduction becomes larger when the half cone angle increases. That is to say, the stiffness softening is more significant when the cross-section area changes more rapidly in the height direction of the sample.

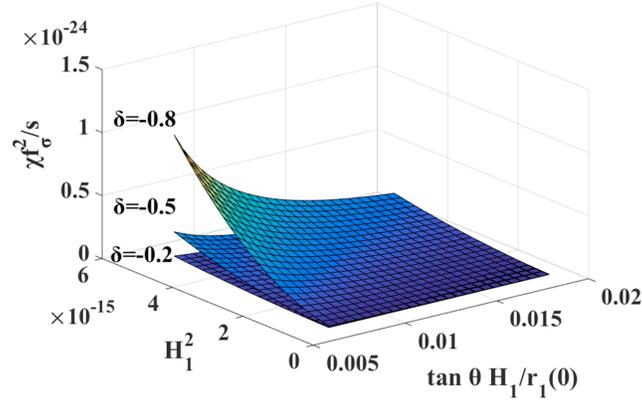

**Figure 5.** The stiffness reduction isosurface that depends on two geometric parameters and one material parameter.

Combining Eqs. (10) and (12), we obtain

$$\frac{1}{\delta} = \frac{C_0}{C_1 - C_0} = \frac{\ln\left(\frac{\tan\theta}{r_1(0)}H_1 + 1\right)}{\frac{\chi f_\sigma^2}{s}\frac{1}{H_1^2}\left(\frac{\tan\theta}{r_1(0)}H_1\right)^2\left[\left(1+\frac{\tan\theta}{r_1(0)}H_1\right)^{-2} - 1\right]} - 1. \qquad (14)$$

It can be seen that the stiffness reduction $\delta$ depends on three parameters, i.e. the geometric parameters $\frac{\tan\theta}{r_1(0)}H_1$ and $H_1^2$, and the material parameter $\chi f_\sigma^2/s$. Using Eq. (14), we can predict the stiffness reduction of the cone-frustum sample with known dimensions and material properties, or design the sample's dimensions for the desired stiffness reduction based on the known or estimated the material properties.

For each given stiffness reduction $\delta$, the three parameters ($\frac{\tan\theta}{r_1(0)}H_1, H_1^2$ and $\chi f_\sigma^2/s$) that satisfy Eq. (14) can make up stiffness reduction isosurface as shown in Fig. 5. It can be seen that the stiffness reduction amplitude becomes larger (from -0.2 to -0.8) when the material parameter $\chi f_\sigma^2/s$ increases and it also becomes larger when the geometric parameter $H_1^2$ decreases or the other geometric parameter $\frac{\tan\theta}{r_1(0)}H_1$ increases. Therefore, we can choose the materials with large $\chi f_\sigma^2/s$ and the samples with small $H_1^2$ and large $\frac{\tan\theta}{r_1(0)}H_1$ in the experimental design process to have better resolution in the measurement. If nanopillars of several nanometers are designed and measured in this method, the expected sensitivity for measuring flexoelectric parameters is about 0.05 μC/m, which is smaller than most of the perovskite ceramics measured by traditional method (i.e. from 0.5 to 150 μC/m).[3] Therefore, the method presented in this work possess enough sensitivity to be applied to measure the nanoscale flexoelectricity with values in the range that aroused scientists' interests in recent years. Moreover, this method avoids measuring the electric polarization or current at nanoscale with dynamical loading, which will simplify the setup of flexoelectric measurement.

In summary, an analytical method is presented for measuring the flexoelectric coupling coefficient of dielectric materials at nanoscale. This method is based on the nanocompression measurement of two samples with different shapes, i.e. one with constant cross-section and the other with variable cross-section such as nano-column and nano-cone-frustum. The flexoelectricity induced reduction in stiffness of nano-cone-frustum is predicted, which becomes more significant when the size of the sample decreases or the half cone angle increases. Two geometric parameters and one material parameter are found to govern the stiffness reduction of the samples. The flexoelectric coupling coefficient can be determined by the mechanical measurement of the stiffness reduction, without electronic polarization or current measurement. This can be beneficial to the flexoelectric measurement at nanoscale and design of nanodevices.

**ACKNOWLEDGMENTS**

The authors are grateful for the support by the National Natural Science Foundation of China (Nos. 11572040), the National Programs for Scientific Instruments Research and Development of China (No. 2012YQ03007502) and the Beijing NOVA Program (No. Z151100000315041).